\documentclass[lettersize,journal]{IEEEtran}
\IEEEoverridecommandlockouts
\usepackage{algorithmic}
\usepackage{amsmath,amssymb,amsfonts}
\usepackage{array}
\usepackage[caption=false,font=normalsize,labelfont=sf,textfont=sf]{subfig}
\usepackage{textcomp}
\usepackage{stfloats}
\usepackage{url}
\usepackage{verbatim}
\usepackage{graphicx}
\usepackage{comment}
\usepackage{booktabs}
\usepackage{cite}
\usepackage{afterpage}
\usepackage{subfig}
\usepackage{listings}
\usepackage{multirow}
\usepackage{tcolorbox}
\usepackage{xspace}
\usepackage{soul}
\usepackage{lipsum}
\usepackage{textcomp}
\usepackage{url}
\usepackage{multirow}
\usepackage[ruled,vlined,linesnumbered]{algorithm2e}
\usepackage{makecell}
\usepackage{rotating}
\usepackage{flushend}
\usepackage{tikz}
\usepackage{bm}
\usepackage[inline]{enumitem}   
\setlist[itemize]{leftmargin=*}
\usepackage{adjustbox}

\newcommand{\pk}{\mathsf{pk}}
\newcommand{\e}{\mathsf{e}}
\newcommand{\m}{\mathsf{m}}
\newcommand{\ct}{\mathsf{ct}}
\newcommand{\s}{\mathsf{s}}

\newcommand{\calB}{\mathcal{B}}
\renewcommand{\c}{\mathsf{c}}

\newcommand{\ring}{\mathcal{R}}

\newcommand{\race}{RACE\xspace}
\newcommand{\rise}{RISE\xspace}
\newcommand{\hecomputing}{HE computing\xspace}
\newcommand{\baseline}{baseline\xspace}
\newcommand{\ciphertexttomessage}{ciphertext-to-message\xspace}
\newcommand{\messagetociphertext}{message-to-ciphertext\xspace}

\newcommand{\ignore}[1]{}

\SetKwComment{Comment}{/* }{ */}
\SetAlFnt{\small}

\usepackage{multirow}
\usepackage{makecell}
\usepackage{multirow}
\usepackage{makecell}
\usepackage{rotating}
\usepackage{flushend}

\SetKwComment{Comment}{/* }{ */}
\SetAlFnt{\small}

\newcommand{\Mod}[1]{\ (\mathrm{mod}\ #1)}

\def\BibTeX{{\rm B\kern-.05em{\sc i\kern-.025em b}\kern-.08em
    T\kern-.1667em\lower.7ex\hbox{E}\kern-.125emX}}

\begin{document}

 \title{RISE: RISC-V SoC for En/decryption Acceleration on the Edge for Homomorphic Encryption}

\author{
     \IEEEauthorblockN{Zahra Azad, Guowei Yang, Rashmi Agrawal, \textit{Student Member, IEEE}, Daniel Petrisko, \textit{Student Member, IEEE},\\ Michael Taylor, \textit{Senior Member, IEEE}, Ajay Joshi, \textit{Senior Member, IEEE}}
     \thanks{Zahra Azad, Guowei Yang, Rashmi Agrawal, and Ajay Joshi are with Boston University, MA, USA (e-mail: \{zazad, guoweiy, rashmi23, joshi\}@bu.edu). \\Daniel Petrisko and Michael Taylor are with University of Washington, WA, USA (e-mail: \{petrisko, profmbt\}@cs.washington.edu)}
 }



\maketitle

\begin{abstract}
Today edge devices commonly connect to the cloud to use its storage and compute capabilities.
This leads to security and privacy concerns about user data.
Homomorphic Encryption (HE) is a promising solution to address the data privacy problem as it allows arbitrarily complex computations on encrypted data without ever needing to decrypt it.
While there has been a lot of work on accelerating HE computations in the cloud, little attention has been paid to the \messagetociphertext and \ciphertexttomessage conversion operations on the edge.
In this work, we profile the edge-side conversion operations, and our analysis shows that during conversion error sampling, encryption, and decryption operations are the bottlenecks.
To overcome these bottlenecks, we present \rise, an area and energy-efficient RISC-V SoC.
\rise leverages an efficient and lightweight pseudo-random number generator core and combines it with fast sampling techniques to accelerate the error sampling operations. 
To accelerate the encryption and decryption operations, \rise uses scalable, data-level parallelism to implement the number theoretic transform operation, the main bottleneck within the encryption and decryption operations.
In addition, \rise saves area by implementing a unified en/decryption datapath, and efficiently exploits techniques like memory reuse and data reordering to utilize a minimal amount of on-chip memory. 
We evaluate \rise using a complete RTL design containing a RISC-V processor interfaced with our accelerator.
Our analysis reveals that for \messagetociphertext conversion and \ciphertexttomessage conversion, using \rise leads up to $6191.19 \times$ and $2481.44 \times$ more energy-efficient solution, respectively, than when using just the RISC-V processor.


\end{abstract}
\section{Introduction}
\label{sec:intro}
Cloud computing has enabled reliable and affordable access to shared computing resources at scale. 
Hence, energy and area-constrained edge devices outsource their computing needs to a third-party cloud system.
However, outsourcing data to a third-party cloud raises data security and privacy concerns.
While an edge device can encrypt the data and send it to the cloud, within the cloud the data needs to be decrypted before processing.
The cloud needs to decrypt the data, which again leaves the data vulnerable to all kinds of data breaches. 

\begin{figure}
\centering
\includegraphics[width=\linewidth]{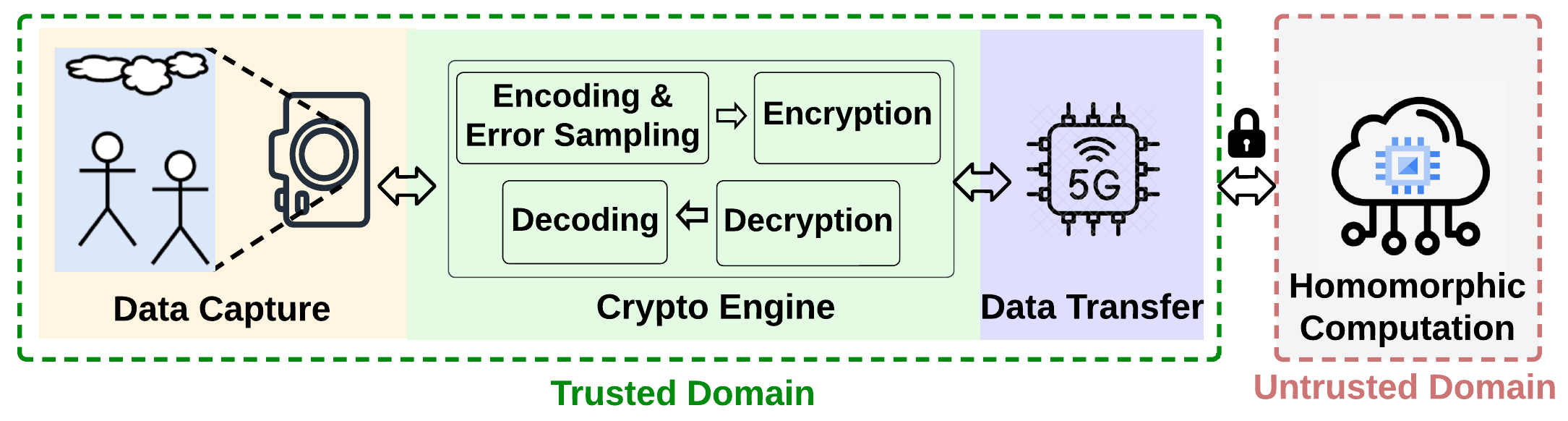}
    \caption{\textit{The dataflow of an end-to-end encrypted computation based on Homomorphic Encryption.}}
    \label{fig:camera}
    \vspace{-0.15in}
\end{figure}

Homomorphic Encryption (HE)~\cite{rivest1978data, gentry2009fully} has emerged as a class of encryption schemes that address this problem by enabling computation on encrypted data. 
Figure~\ref{fig:camera} shows an illustrative use case of how HE can be used to outsource secure computation.
A user captures an image/video using an edge device.
The captured image/video is encoded and encrypted on the edge device and then transferred to a third-party cloud system. 
The untrusted cloud system can process the encrypted data and send the encrypted result back that can be decrypted and decoded only by the user.

Although HE-based privacy-preserving computing seems plausible, it is several orders of magnitude slower than operating on unencrypted data~\cite{natarajan2021seal}. 
To bridge this performance gap, several existing works take advantage of software and hardware optimizations to accelerate cloud-side HE operations running on CPU~\cite{he_cloud_6, he_cloud_8}, GPU~\cite{he_cloud_1,9201530}, and custom hardware accelerators~\cite{he_cloud_2, he_cloud_3, he_cloud_4, he_cloud_5, he_cloud_7}. 
Unfortunately, little attention has been paid to edge-side operations even when the edge-side operations are non-trivial.
For encrypting the data, the edge device needs to perform encoding, error sampling, and encryption.
These three operations together form the ``message-to-ciphertext'' conversion operation.
Similarly for decrypting the data received from the cloud, the edge device needs to perform decryption and decoding.
These two operations together form the ``ciphertext-to-message'' conversion operation.
These edge-side operations incur huge memory consumption (on the order of several MBs) and computation overhead.

To accelerate these edge-side operations, recently Microsoft released SEAL-Embedded~\cite{natarajan2021seal} as the first HE library targeting embedded devices. 
SEAL-Embedded proposes a number of optimizations for error sampling, en/decoding, and en/decryption on resource-constrained edge devices. 
To enable computing on a variety of data captured by the sensors on the edge device, SEAL-Embedded targets Cheon-Kim-Kim-Song (CKKS)~\cite{CKKS17} HE scheme as it enables operations on real numbers. 
Unfortunately, this implementation of the library is still not practical.
For example, the industry-required frame rate for surveillance cameras and mobile platforms typically ranges from $15$ to $60$ frames per second~\cite{usman2018intrusion}.
With the SEAL-Embedded library running at $1$ GHz on a RISC-V processor like BlackParrot~\cite{blackparrot}, for a polynomial of degree $N = 4096$ and three $30$-bit primes\footnote{These are the largest parameters supported by the SEAL-Embedded library.}, we are unable to encrypt even a single low resolution quarter quarter video graphics array (QQVGA) frame per second (further details in Section~\ref{sec:evaluation}).
While one could use a more powerful processor to achieve the required frame rates, it comes at the cost of high power consumption, which is not acceptable in edge devices.

\begin{figure*}[t]
\centering
\includegraphics[width=\linewidth]{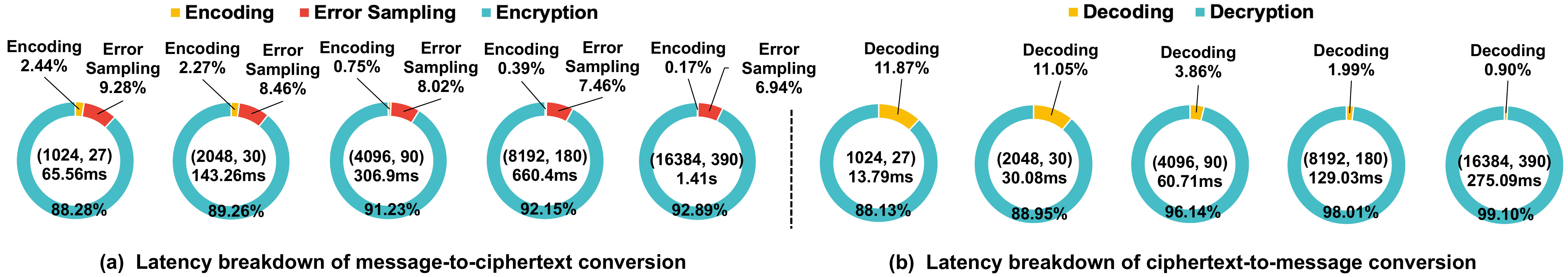}
    \caption{\textit{Latency breakdown of (a) \messagetociphertext and (b) \ciphertexttomessage conversion operations running on BlackParrot using SEAL-Embedded library. Corresponding scheme parameters ($N$, $\log Q$) and latencies are specified inside the doughnut charts.}}
    \label{fig:breakdown_sw}
\end{figure*}

As software solutions are inefficient, several prior works focused on accelerating the key performance bottlenecks within the edge-side operations in hardware. 
Figure~\ref{fig:breakdown_sw} (a) and (b) shows the latency breakdown of the \messagetociphertext and \ciphertexttomessage conversion for different scheme parameters (polynomial degree, $N$ and coefficient bit-width, $\log Q$) running on BlackParrot using SEAL-Embedded library. 
For all the parameter sets that we evaluated, the encryption and decryption operations incur the highest latency because they perform multiple polynomial multiplications.
The latency of the encryption and decryption operations is dominated by number theoretic transform (NTT) operations. 
Error sampling is also a bottleneck operation accounting for up to $10$\% of the total \messagetociphertext conversion latency.

To address these performance bottlenecks, there exist works that focus on accelerating sub-operations like NTT~\cite{ nannipieri2021risc, li2021high, chen2022cfntt, duong2021configurable, replace1, banerjee2019sapphire, Roy2014, ye2022pipentt, paludo2022ntt, su2022highly, duong2022area} in en/decryption and pseudo-random number generation (PRNG)~\cite{nannipieri2021risc, li2021high, chen2022cfntt} in error sampling.
For accelerating the complete encryption and decryption operations, Su et al.~\cite{en_he_fpga} and Yoon et al.~\cite{yoon201955nm} proposed an FPGA-based and an ASIC-based accelerator, respectively, targeting Brakerski-Gentry-Vaikuntanathan (BGV) HE scheme~\cite{gentry2012ring}. 
Both of these solutions use small scheme parameters ($N<=2^{10}$ and $\log Q<=24$).
However, these parameters are not practical for most real-world applications because HE schemes contain a noise term (error sample) within the ciphertext coefficients, which is essential for security.
This noise within the ciphertext increases with each succeeding homomorphic operation until it reaches a critical level at which it is impossible to recover the computation output~\cite{gentry2009fully}. 
To increase the noise budget for practical HE applications with a large number of HE computations, we need large scheme parameters ($N>2^{12}$ and $\log Q>109$).

In this work, we present \rise, a System-on-Chip (SoC) containing a RISC-V BlackParrot core and an area and energy-efficient hardware accelerator that supports large scheme parameters ($N$ and $\log Q$), which enables practical CKKS-based HE applications.  
To address the performance bottlenecks, in \rise we accelerate the error sampling and en/decryption operations while reducing the area overhead and energy consumption. 
To speed up error sampling, we take an efficient and lightweight implementation of a PRNG core~\cite{WinNT} and integrate it with fast binomial and uniform samplers. 
We propose a shared datapath (referred to as a unified datapath later in the paper) for the encryption and decryption operations because they both involve similar operations (polynomial addition and multiplication). 
To reduce on-chip memory (designed using SRAM) area, we manage the data in \rise such that it does not require memory larger than what is required to store two polynomials. 
In contrast to prior works~\cite{nannipieri2021risc, li2021high, chen2022cfntt, duong2021configurable, replace1, mert2020flexible} that require dual port ($1$R$1$W) SRAM banks to access polynomial coefficients for NTT computation, we propose a novel data reordering scheme for NTT so that \rise only needs single port ($1$RW) SRAM banks, which further reduces the area. 
Moreover, \rise exploits the data-level parallelism in NTT by leveraging a scalable parallel implementation of butterfly operations.

The main contributions of our work are as follows:
\begin{itemize}
    \item We profile edge-side operations for the CKKS scheme by executing the SEAL-Embedded library on BlackParrot core (referred to as \baseline in the rest of the paper), for a range of scheme parameters ($N$ and $\log Q$) to identify the performance bottlenecks.
    \item Based on the profiling results, we architect \rise, an area- and energy-efficient SoC (containing BlackParrot core and an accelerator) to accelerate the error sampling and en/decryption operations. 
    We use several optimizations such as data level parallelism, a shared data path for the en/decryption operations, memory reuse and data reorder techniques to architect an efficient accelerator design. 
    \item  We evaluate \rise by executing \messagetociphertext and \ciphertexttomessage conversion operations using performance, area, and energy efficiency metrics.
    Across a range of parameters, \rise reduces the \messagetociphertext and \ciphertexttomessage conversion latency by $28.79\times$-$104.39 \times$ and $7.95 \times$-$66.08 \times$, respectively, as compared to the \baseline.
    \rise achieves $471.24 \times$-$6191.19 \times$ lower EDP when performing \messagetociphertext conversion and $36 \times$-$2481.44 \times$ lower EDP when performing \ciphertexttomessage conversion as compared to \baseline.
    Similarly, \rise has $24.06 \times$-$55.36 \times$ lower ADP when performing \messagetociphertext conversion and $6.65 \times$-$35.05 \times$ lower ADP when performing \ciphertexttomessage conversion as compared to \baseline. 
\end{itemize}

\section{Preliminaries}
\label{sec:prelim}

\subsection{Homomorphic Encryption}
\label{subsec:he-RLWE}
The \hecomputing model allows operating on encrypted data to maintain data privacy. 
Over the years, a variety of HE schemes have been developed such as BGV~\cite{gentry2012ring}, Brakerski/Fan-Vercauteren (BFV)~\cite{brakerski2012fully}, and CKKS~\cite{CKKS17}.
The CKKS scheme allows operations on real numbers, which are required for various applications including machine learning, scientific, and graph applications.
Hence, we choose to focus on the CKKS scheme in our paper, and we use SEAL-Embedded library to implement it. 
Below we describe the process for \messagetociphertext and \ciphertexttomessage conversion.

The CKKS scheme works with a native plaintext data type that is a vector of length $N/2$, where each element is chosen from the field of complex numbers $\mathbb{C}$. 
The encoding operation takes as input this $N/2$-dimensional vector and returns polynomial $\m(X)$ with integer coefficients.
The polynomial $\m(X)$ can be encrypted under the public key $\pk$, generating a ciphertext $\ct$ by computing: 
\begin{align}
\label{eq:enc1}
    \c_0 = \mu \cdot \pk_0 + \m + \e_0, \\
\label{eq:enc2}
    \c_1 = \mu \cdot \pk_1 + \e_1 
\end{align} 

Here, the $\mu$ polynomial is sampled from a uniform distribution, and the error polynomials $\e_0$ and $\e_1$ are sampled using a discrete Gaussian noise sampler.
The coefficients in the ciphertext polynomials ($\c_0, \c_1$) are elements of $\mathbb{Z}_Q$, where $\mathbb{Z}$ is a set of integers and $Q$ defines the order of finite field.
Here modulus $Q$ is typically on the order of thousands of bits to account for the noise growth.
The CKKS scheme supports the use of Residue Number System (RNS) (also known as the Chinese Remainder Theorem (CRT) representation) to compute on such large operands efficiently. 
Using the RNS approach, each coefficient is represented modulo $Q = \prod_{i=1}^\ell q_i$, where each $q_i$ is a prime number.
We can represent $x \in \mathbb{Z}_Q$ as a length-$\ell$ vector of scalars $[x]_\calB = (x_1, x_2, \ldots, x_\ell)$, where $x_i \equiv x \pmod{q_i}$. 
We refer to each $x_i$ as a \emph{limb} of $x$.
The ciphertext is decrypted to obtain the original message back using the following equations: 
\begin{align}
\label{eq:dec}
    \m = \c_0 + \c_1 \cdot \s \Mod {q_{\ell}} 
\end{align} 
Here $\s$ is the secret key.
Using RNS, both encryption and decryption can be performed w.r.t. a smaller modulus $q_i$ instead of a large modulus $Q$.

\ignore{
\noindent \textbf{Video Frame Encryption Example:}
Considering our example of video frame encryption, using Quarter Quarter VGA (QQVGA) frame resolution, the frame size is $120 \times 160$ pixels.
If this frame is in grayscale, the frame size will be $120 \times 160 \times 8 = 153,600$~bits = $19.2$~KB.
With $N = 4096$ and $\log q = 30$ bits, we can encode $N/2 \times \log q = 2048 \times 30 = 61, 440$ bits in a single ciphertext, which implies that a single frame will be encoded and encrypted within $3$ ciphertexts and will have a total size of $327$~KB.
}

\subsection{Number Theoretic Transform}
\label{subsec:ntt}
Polynomial multiplication is a critical step in encryption and decryption operations.
A na\"ive approach to perform a polynomial multiplication has a complexity of $O(N^2)$ multiplications for a polynomial of degree $N$.
Therefore, to reduce this computational complexity, an NTT operation is applied to the polynomials so as to perform a point-wise multiplication.
Using NTT we can reduce the polynomial multiplication complexity to $O(N\log{}N)$. 
NTT can be viewed as the finite field version of fast Fourier Transform (FFT).
During NTT, coefficients of the input polynomial are multiplied with the power of an $N$-th primitive root of unity and combined with each other in a butterfly fashion.
Before each polynomial multiplication takes place in an encryption operation (see Equation~\eqref{eq:enc1} and~\eqref{eq:enc2}), the polynomials are converted into an NTT domain.
Similarly, we need to perform an inverse NTT (iNTT) operation in the decryption operation.
Both NTT and iNTT operations add high computational complexity to the encryption and decryption operations, respectively.

\subsection{BlackParrot: RISC-V Multicore Processor}
\label{subsec:bp}
BlackParrot is an agile open-source RISC-V multicore processor for accelerator SoCs~\cite{blackparrot}.
The BlackParrot multicore implements the RISC-V $RV64G$ architecture and is designed as a scalable, heterogeneously tiled multicore microarchitecture.
BlackParrot microarchitecture has four different tile types: 
\begin{enumerate*} 
    \item A Core Tile, which contains a BlackParrot processor with one or more coherent caches, a directory shard, and an L2 slice, 
    \item An L2 extension tile, which is used to scale-out the on-chip L$2$ in the BlackParrot system, 
    \item A Coherent accelerator tile, which has a local cache engine (LCE) with a backing coherent cache, and 
    \item A Streaming accelerator tile, which does not have a cache memory behind their LCE link and does not control any physical memory. Streaming tiles can be used for basic I/O devices, network interface links, or GPUs. 
\end{enumerate*} 

BlackParrot provides a robust and scalable end-to-end framework for accelerator integration, which simplifies the interfacing of both coherent and streaming accelerators, and the offloading of parts of the user application from the processor to the accelerator.
This framework provides hardware implementation of streaming and coherent accelerator tiles in SystemVerilog (simulation and FPGA prototype).
This helps accelerator designers and system architects to evaluate their accelerator related ideas using hardware implementation rather than simulation, and find the integration strategy that has low offload and synchronization overheads for their application to improve the end-to-end application time. 

\subsection{Video Encryption Example}
\label{subsec:video_encryption}
In this paper, we use the example of video encryption to discuss the choice of $N$ and $\log Q$, sizes of the ciphertext, memory and compute requirements for \messagetociphertext and \ciphertexttomessage conversions, and how that influenced the microarchitecture of \rise.
A video is made up of multiple frames, where a frame size is defined by $f_w \times f_h \times b_{pp}$.
Here, $b_{pp}$ defines the bits per pixel and assumes a value of $8$ for a grayscale pixel.
For a given $N$, $\log q$, and $limbs$ value, we can encode $N/2 \times \log q$ bits in a single ciphertext, which implies that a single frame will be encoded and encrypted within multiple ciphertexts (cts) and will have a total size of $N \times \log q \times limbs \times \#cts$ bits.

For a QQVGA, the frame resolution is $120 \times 160$ pixels.
If this frame is in grayscale, the frame size will be $120 \times 160 \times 8 = 153,600$~bits = $18.75$~KB.
With $N = 4096$ and $\log q = 30$ bits, we can encode $N/2 \times \log q = 2048 \times 30 = 61, 440$ bits in a single ciphertext, which implies that a single frame will be encoded and encrypted within $3$ ciphertexts and will have a total size of $270$~KB.
While $N=4096$ and $\log Q=90$ bits parameter set provides $128$-bit security, to enable practical applications using \hecomputing approach, we need to have larger parameters such as $N=16384$ and $\log Q=390$ bits.
For this $N$ and $\log Q$ combination, a single frame will be encoded and encrypted within a single ciphertext and will have a total size of $1.6$~MB.
Similarly for QVGA, the frame resolution is $320 \times 240$ pixels.
If this frame is in grayscale, the frame size will be $320 \times 240 \times 8 = 614,400$~bits = $75$~KB.
With $N = 4096$ and $\log q = 30$ bits, we can convert $N/2 \times \log q = 2048 \times 30 = 61, 440$ bits of a frame in a single ciphertext, which implies that a single frame will be encoded and encrypted within $10$ ciphertexts and will have a total size of $900$~KB.
With $N = 16384$ and $\log q = 30$ bits, a single frame will be encoded and encrypted within a single ciphertext and will have a total size of $4.5$~MB.

Given the limited on-chip memory in edge devices, we cannot use batch processing for \messagetociphertext and \ciphertexttomessage conversion of the frames.
We need to architect \rise such that it can match the throughput of the \messagetociphertext conversions with the typical frame rates of $15$ to $60$ frames per second.
In contrast, the \ciphertexttomessage conversion is constrained by the bandwidth ($100-900$ Mbps) of the network connecting the edge device and the cloud.
\section{Related Work}
\label{sec:rel-work}
Over the years, there have been several works that have focused on accelerating \hecomputing on the cloud side.
These works include algorithmic optimizations for CPU~\cite{he_cloud_6, he_cloud_8} and GPU~\cite{he_cloud_1,9201530}, and custom hardware accelerators~\cite{he_cloud_2, he_cloud_3, he_cloud_4, he_cloud_5, he_cloud_7} running in the cloud.
All these works assume that the cloud receives encrypted data from the edge device and that the cloud sends the encrypted result back to the edge device for decryption.
There is an implicit assumption in these works that the edge devices have the capability to encrypt and decrypt the data with high performance and do not need any hardware acceleration.
However, the encryption and decryption of data for \hecomputing is compute intensive and has a very high memory usage.
For the edge devices that are constrained by power, performance, and area, we need to develop an efficient solution for edge-side operations.

\subsection{Software-based Solutions} 
\label{subsec:sws}
Microsoft SEAL~\cite{seal} is a HE library that allows addition and multiplication operations on encrypted integers or real numbers. 
Recently, SEAL has been extended to SEAL-Embedded~\cite{natarajan2021seal} for resource-constrained edge devices.
SEAL-Embedded exploits RNS partitioning, data type compression, memory pooling, and reuse to reduce the memory consumption.
However, this software-based implementation of encryption operation is still slow and not efficient for real-time applications. 
As mentioned earlier, for a video application with a low resolution of QQVGA, SEAL-Embedded fails to encrypt even one frame per second running at $1$~GHz on a RISC-V core like BlackParrot~\cite{blackparrot} for a practical set of scheme parameters (polynomial degree of $N = 4096$ and three $30$-bit primes). 

\subsection{Hardware-based Solutions}
\label{subsec:hws}
There are a few works focusing on accelerating edge side operations for HE~\cite{en_he_fpga, yoon201955nm}. 
Su et al. ~\cite{en_he_fpga} present an FPGA-based accelerator for the BGV HE scheme as against the CKKS scheme that we support. 
Their BGV accelerator only supports small scheme parameters ($N = 128$, $\log Q=27$), which are impractical for HE computation.
However, the authors claim that their accelerator can be extended to larger polynomial degrees to support higher security levels, but support for larger parameters is left as future work. 
Moreover, the accelerator is mainly optimized to achieve high performance and throughput, while ignoring area/energy efficiency. 
Yoon et al.~\cite{yoon201955nm} present an ASIC-based en/decryption accelerator for HE operations. 
The accelerator is again evaluated only for small parameters ($N=16$). 
Even to support these small polynomials, it needs large buffers to store the in/outputs and the pre-computed twiddle factors, increasing the memory area. 

In our work, we architect an accelerator that can perform \messagetociphertext and \ciphertexttomessage conversions for practical scheme parameters.
Our accelerator uses data-level parallelism, shares the datapath between encryption and decryption operations, adopts memory reuse and memory reordering strategies, and eliminates the need for additional on-chip memory to store twiddle factors by computing them on-the-fly. 
\section{\rise System View}
\label{sec:systemview}
\begin{figure*}[t]
\centering
\includegraphics[width=\linewidth]{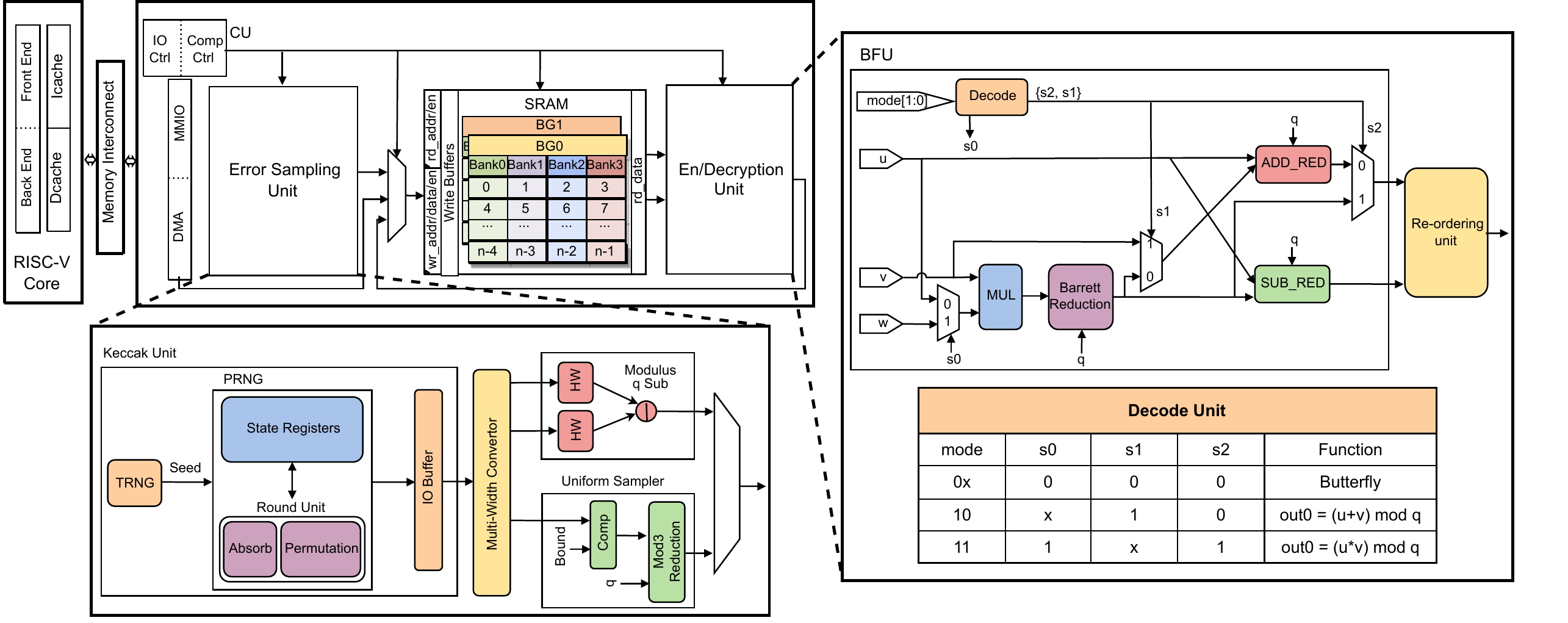}
    \caption{\textit{System-level view of \rise, a RISC-V SoC for accelerating \messagetociphertext and \ciphertexttomessage conversion operations on the edge for supporting homomorphic operations in the cloud.}}
    \label{fig:he_bp}
\end{figure*}
In this section, we present the overall design of \rise, an end-to-end SoC (see Figure~\ref{fig:he_bp}) that consists of a single BlackParrot RISC-V core, and an accelerator that performs error sampling, encryption, and decryption.
The accelerator is interfaced with the BlackParrot core in a streaming fashion because a large amount of data needs to be frequently transferred between the two.
To move all the input data from the main memory of BlackParrot core to the accelerator, we configure a hardware DMA logic.
The user provides public keys ($\pk_0$, $\pk_1$) and input message $\m$ to the BlackParrot core.
The BlackParrot core is responsible for performing en/decoding operations and the random seed generation using SEAL-Embedded library.
The PRNG unit in accelerator receives the random seed from the BlackParrot core and uses it to generate a bit stream of pseudo-random numbers. 
These pseudo-random numbers are passed to a fast error sampler to generate the required error polynomials, i.e., $\e_0$, $\e_1$, and $\mu$.
These error polynomials along with the encoded message and public keys are then used to perform encryption.
The encryption operation performs the operations described in the Equations~\eqref{eq:enc1} and \eqref{eq:enc2}.
Similarly, the decryption operation performs the operations listed in Equation~\eqref{eq:dec}.
For decryption operation, we need the ciphertext ($\c_0$ and $\c_1$) that is sent by the cloud and the secret key that is generated by the BlackParrot core as inputs.
Once the en/decryption operation is completed, the BlackParrot core receives an interrupt from the accelerator.
Then, the DMA logic transfers the output of the accelerator back to the main memory of the BlackParrot core.

\section{Accelerator Microarchitecture}
\label{sec:microarch}
In this section, we provide a detailed description of the microarchitecture of our accelerator (see Figure~\ref{fig:he_bp}).

\subsection{Error Sampling Unit}
\label{subsec:error}
Error samples are critical to maintaining the required security level while performing HE operations. 
However, generating these high-quality error samples is one of the bottlenecks in the edge-side operations.
As shown in Figure~\ref{fig:he_bp}, error sampling basically consists of two steps: generation of pseudo-random numbers using a true random seed, and generation of uniform and binomially distributed error samples using the generated pseudo-random numbers.
Below we present the microarchitecture of a lightweight PRNG, a binomial sampler, and a uniform sampler.\\

\noindent \textbf{Pseudo-Random Number Generator (PRNG):} 
We have a customized PRNG unit as part of the accelerator to speed up pseudo-random number generation process~\cite{WinNT}. 
One of the prior works~\cite{banerjee2019sapphire} evaluated various PRNGs and concluded that the SHA-$3$ hash family in the SHAKE mode~\cite{dworkin2015sha}, is $2\times$ and $3\times$ more energy efficient than ChaCha$20$~\cite{bernstein2008chacha} and AES~\cite{dworkin2001advanced}, respectively.
This is due to the fact that SHA-$3$ in SHAKE mode generates the highest number of pseudo-random numbers per round. 
Therefore, in our PRNG unit design, we use a SHAKE function, which is more commonly referred to as Keccak. 
For our use case of en/decryption operation that requires a large number of error samples (as $N$ is ${>}2^{12}$), Keccak makes a perfect PRNG because its output length is not predetermined.
Hence, we can generate as many error samples as needed for the en/decryption operation with just one invocation of the Keccak unit. 

A Keccak unit typically consists of a round unit with two sub-units: Absorb and Permutation.
A true random seed (generated by TRNG), and the desired length of the pseudo-random number, and the rate at which the pseudo-random numbers are generated (provided by the BlackParrot core via control and status registers (CSRs)) are input to the Absorb sub-unit. 
In our design, the true random seed consists of $1600$ bits.
A Keccak round operates on the data organized as an array of $5 \times 5$ computation lanes, each of length $64$. 
Hence, the absorption phase changes the random seed from a $1$D $1600$-bit representation into a $2$D $25 \times 64$-bit representation, and we store this $2$D representation in a state register (see Figure~\ref{fig:he_bp}). 
The value in the state register is permuted by performing a series of shift, XOR, AND, and NOT operations in the Permutation unit~\cite{WinNT}. 
We store the output of the Permutation unit in the state register. 
We set the length of the pseudo-random number to $1088$ bits, which is the maximum length supported by Keccak. 

     
\noindent \textbf{Error Sampler:} The output of PRNG is passed to a uniform sampler and a binomial sampler to generate error polynomials. 
For RLWE cryptosystems, the original worst-case to average-case security reductions hold for both continuous (rounded) Gaussian distributions and discrete Gaussian distributions. 
However, the implementation of efficient and constant-time Gaussian sampling is a challenging problem~\cite{agrawal2020post}. 
Prior works~\cite{natarajan2021seal, xin2020vpqc, duong2021configurable} address this by calculating the difference of the hamming weights of two random bit streams of length $k$, and can rapidly obtain samples from a zero-centered binomial distribution\footnote{Presuming this error distribution's standard deviation is sufficiently large, no known attack exploits the shape of this distribution. For our binomial distribution, we use a standard deviation of $\sqrt{21/2}\approxeq 3.24$ to comply with the HE security standard~\cite{albrecht2021homomorphic}.} in a constant time.
We adopt the same approach in our design.

Additionally, we implement a uniform sampling unit that uniformly samples the coefficients of the polynomial from $\ring_3$ (i.e., $N$ coefficients sampled uniformly from $\{$-1$,0,1\}$). 
We implement this functionality using a rejection sampling algorithm~\cite{alkim2016post}. 
The implementation is a constant time implementation of modulo$3$ reduction (see Figure~\ref{fig:he_bp}).
     

\subsection{Encryption and Decryption Unit}
\label{subsec:encdec}
\begin{figure}[t!]
\centering
\includegraphics[width=\linewidth]{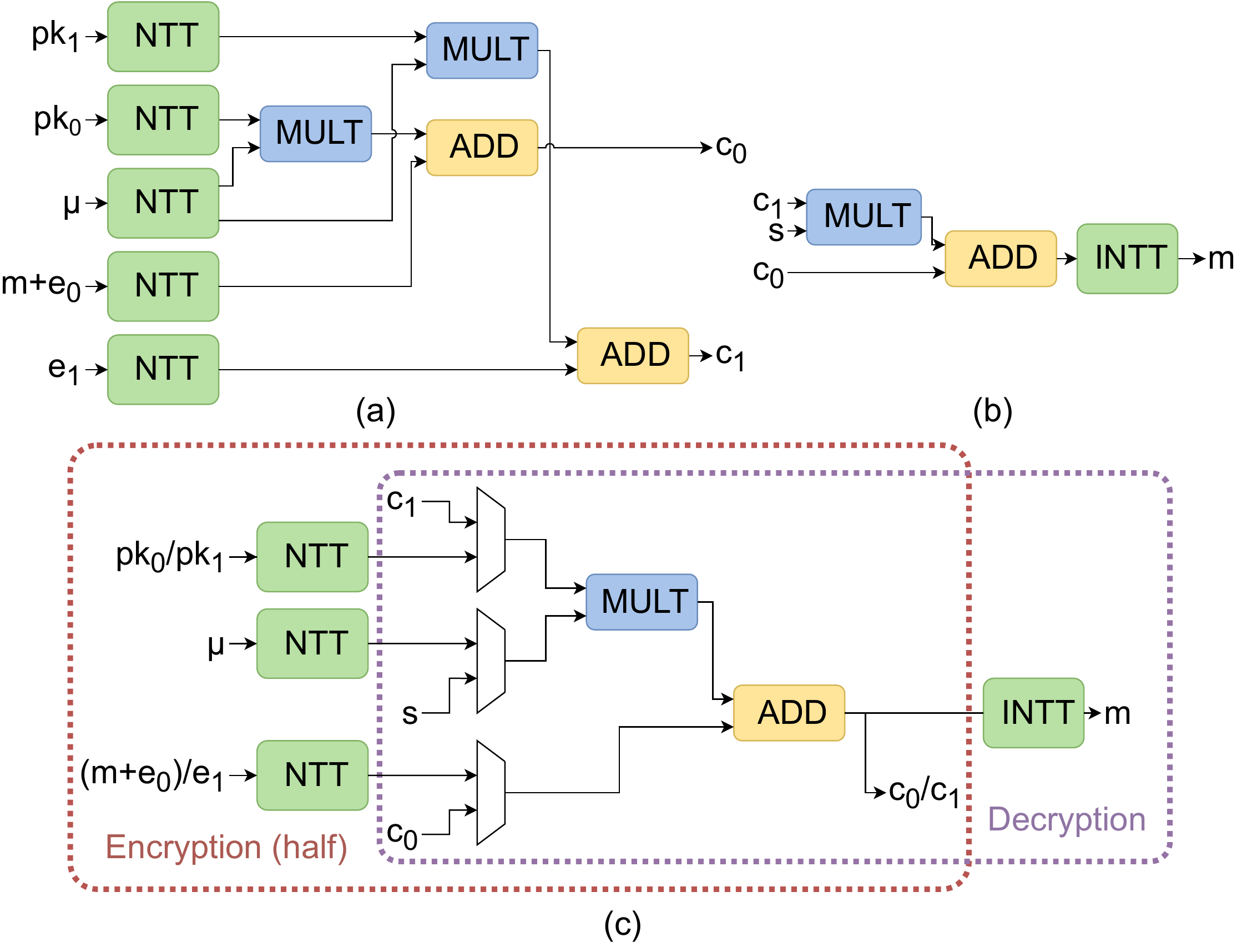}
    \caption{\textit{(a) Encryption dataflow. (b) Decryption dataflow. (c) Unified en/decryption dataflow. In the unified dataflow, encryption and decryption operations share the datapath and the control logic.}}
    \label{fig:accel-dataflow}
\end{figure}
Figure~\ref{fig:accel-dataflow} (a) shows the encryption datapath, which follows Equation~\eqref{eq:enc1} and~\eqref{eq:enc2}.  
Each encryption operation calls the accelerator twice, once to compute $\c_0$  with $(\pk_0, \mu, \m, \e_0)$ input set and then to compute $\c_1$ with $(\pk_1, \mu, \e_1)$ input set. 
The datapath consists of polynomial addition and multiplication operations. 
The polynomial addition involves simple element-wise modular addition of the polynomial coefficients and has a complexity of $O(N)$.
In contrast, polynomial multiplication has a complexity of $O(N^2)$, and like prior efforts, we accelerate it using NTT. (more details about NTT are in Section~\ref{subsec:ntt}).
Acceleration using NTT reduces the complexity of polynomial multiplication to $O(N)$.
Similarly, Figure~\ref{fig:accel-dataflow} (b) shows the datapath for the decryption operation that follows Equation~\eqref{eq:dec}.
Decryption datapath again performs polynomial addition and multiplication operation.
It receives input polynomials that are already in the NTT domain.
However, the decrypted polynomial is required to be in coefficient form for performing the decoding operation (we perform this operation on the BlackParrot core using the SEAL-Embedded library).
Therefore, the decryption datapath has an iNTT operation.\\

\noindent \textbf{Unified En/Decryption Datapath:} 
In order to reduce the area overhead of the accelerator, we share the datapath and control logic of the accelerator between encryption and decryption operations (see Figure~\ref{fig:accel-dataflow} (c)).  
This is possible because the sequence of operations performed in the encryption and decryption operations are the same.
Moreover, the encryption operation uses the exact same sequence of operations to compute both $\c_0$ and $\c_1$.
Thus, we use the same datapath twice to perform the complete encryption operation.\\

\noindent \textbf{NTT Acceleration:} The main performance bottleneck in the en/decryption unit is the NTT operation.
Consequently, we propose several optimization techniques to efficiently perform NTT while incurring a low memory and area overhead. 
We discuss these optimizations in detail in the rest of this section. \\

\noindent \textit{Butterfly Unit (BFU):} 
A Butterfly operation is the basic building block of NTT/iNTT operation.
An NTT/iNTT operation consists of $\log_2N$ stages (for a polynomial of degree $N$), and each stage requires $N/2$ Butterfly operations.
Each BFU takes two coefficients (say $a$ and $b$) out of the $N$ polynomial coefficients as input and computes $(a, b)=(a+\omega \cdot b \Mod{q}$, $a-\omega \cdot b\Mod{q})$ (refer Algorithm~\ref{Algorithm1} line~\ref{algnl:bf1} and \ref{algnl:bf2}).
Here, $\omega$ is the twiddle factor.
A degree $N$ polynomial requires $N/2$ twiddle factors, where each twiddle factor needs $\log q$ bits. 
Our accelerator computes twiddle factors on-the-fly within BFU to reduce the memory overhead for storing them as pre-computed values. 

BFU is fully-pipelined with the throughput of $1$ Butterfly operation per cycle.
It is designed to perform NTT, iNTT, polynomial addition, and multiplication operations that are required by both encryption and decryption operations (see Figure~\ref{fig:accel-dataflow}).
BFU has an integer adder and subtractor unit that performs modular reduction using a conditional operator.
BFU contains a modular multiplier where modular reduction operation is performed using a Barrett reduction~\cite{barrett1986} unit.
Barrett reduction computes modular reduction operation without performing any division and only involves two multiplications and one subtraction, shift, and conditional subtraction operation~\cite{barrett1986}.
In addition, it does not exploit any property of the modulus $q$, which makes it ideal for supporting configurable moduli.
The modular multiplier lies on the critical path in the accelerator.
Hence, we pipeline the multiplier to reduce the critical path and improve the operating frequency of the accelerator.
As power and area are the primary design goals for edge devices, all the above computations are performed by sequentially leveraging the pipelined BFU.\\ 

\begin{figure}[t]
\centering
\includegraphics[width=\linewidth]{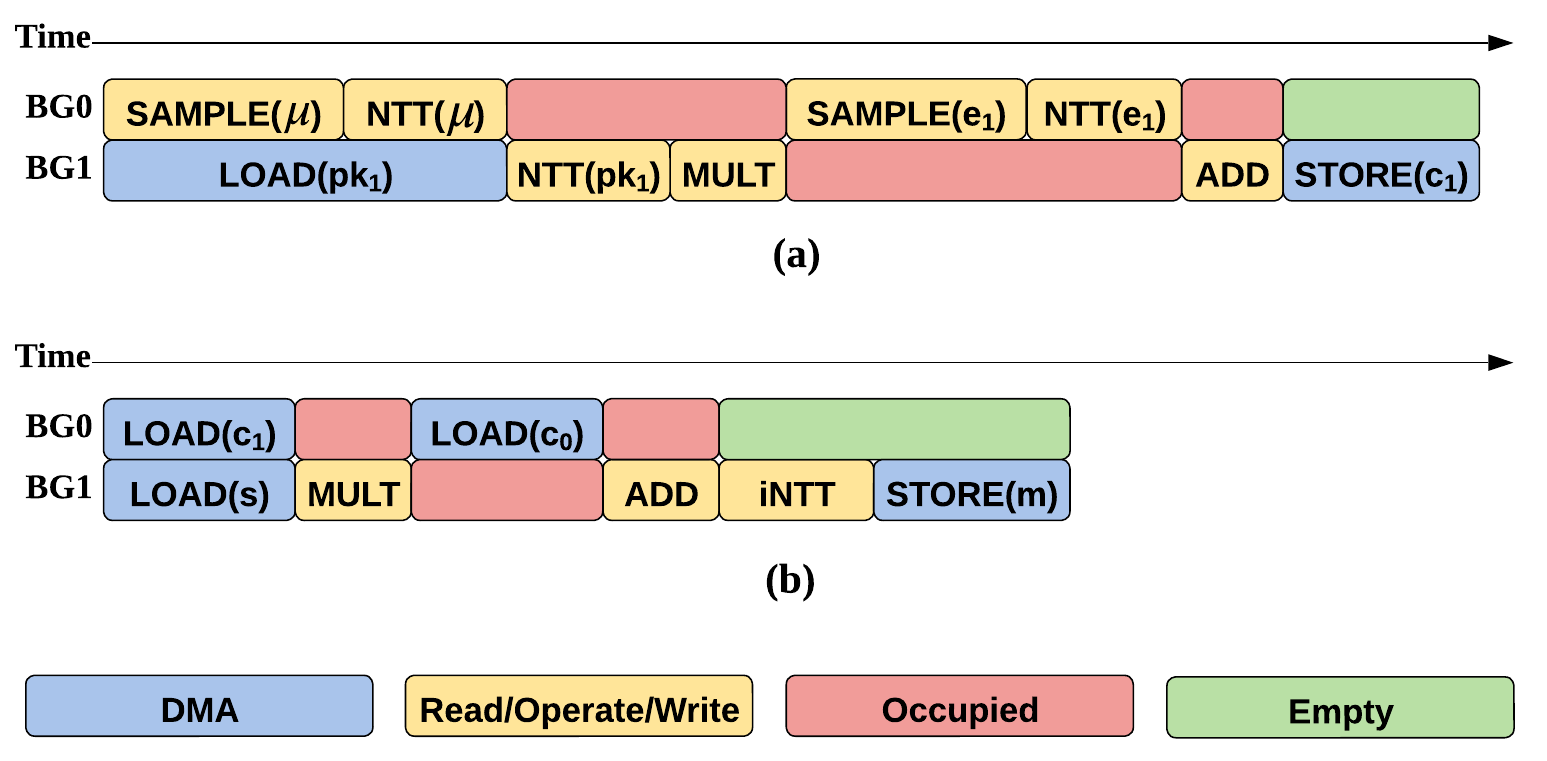}
    \caption{\textit{Memory reuse during (a) encryption and (b) decryption operations. Each BG can store only one polynomial. ``Read\slash Operate\slash Write'' means the bank group is being accessed during the operations. ``Occupied'' means the bank group stores intermediate results.}}
    \label{fig:accelmem}
\end{figure}

\noindent \textit{Memory Reuse Technique:} 
All the necessary polynomials ($\m$, $\e_0$, $\e_1$, $\mu$, $\pk_0$, $\pk_1$, $\c_0$, $\c_1$) should be kept in the accelerator's on-chip memory for efficient en/decryption computation.
A single polynomial typically needs memory of ${\sim}60$~KB with $N=2^{14}$ and $\log q=30$.
Therefore, we require a total of $480$~KB to hold all the in/output polynomials.
In our memory reuse strategy, we manage the encryption and decryption operations such that at any given time, we need to store a maximum of only two polynomials, which takes $120$~KB of space.

For memory reuse, we divide the entire on-chip SRAM memory into multiple banks that are organized into two bank groups, i.e., BG$0$ and BG$1$.
Each bank group corresponds to a single polynomial and each polynomial is stored across multiple banks within a bank group. 
During the encryption and decryption operation, we use these bank groups to store the input, output, and intermediate polynomials. 
Hence, we share each of the two bank groups among several polynomials as shown in Figure~\ref{fig:accelmem} (a) and (b).
As an illustration (see Figure~\ref{fig:accelmem}), we carry out an in-place NTT in an encryption operation that gets the data for polynomial $\mu$ from BG$0$, processes it, and then writes the results back to BG$0$. 
While we are still performing NTT on the polynomial $\mu$, we load the next input polynomial $\pk_1$ into BG$1$ in parallel.
The modular addition and multiplication operations involve memory reuse as well.
Both of these operations read the input from BG$0$ and BG$1$ while writing the output to bank group BG$1$.
Therefore, after the modular addition or multiplication operations are complete, we can reuse BG$0$ for the subsequent operation.
Therefore, by utilizing a memory reuse strategy, we can efficiently perform en/decryption operations while incurring a minimal memory footprint.\\ 

 
\begin{figure*}[t]
\centering
\includegraphics[width=\linewidth]{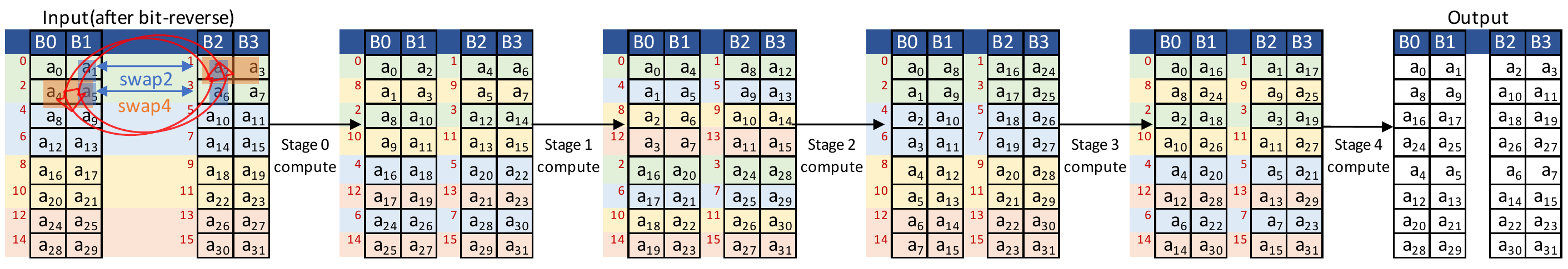}
    \caption{\textit{NTT\_swap$4$ with $N=32$. 
    The red-colored numbers before each pair of cells denote the order of Butterfly operations. 
    The four consecutive Butterfly operations ($2$ rows) being reordered are denoted with the same color.
    }}
\label{fig:accel-swap4-example}
\end{figure*}

\noindent \textit{Memory Reorder Technique:} 
The next memory level optimization that we perform is memory reorder, which helps reduce the number of memory ports required, resulting in low memory area overhead.
Every Butterfly operation takes as input two coefficients of the polynomials, operates on them, and stores the resultant values back to the same memory banks. 
As a result, a na\"ive implementation of NTT will require $2$ read and $2$ write ports ($2$R$2$W) for each memory bank that is of size $N$.
Typically, a $2$R$2$W memory bank is roughly twice as large as $1$ read and $1$ write port ($1$R$1$W) memory bank of the same size.
Consequently, we can save half of the memory area simply by switching from a single $2$R$2$W bank of size $N$ to two $1$R$1$W banks of size $N/2$. 

However, managing memory access patterns for NTT, with $1$R$1$W banks, becomes challenging as the memory accesses can lead to bank conflicts.
Throughout all the NTT stages, the distance ($(j-i)$) between the two inputs of a Butterfly operation changes. 
This leads to bank conflicts in several stages of NTT as each stage in NTT iterates through all values from $1$ to $N/2$.
Thus, replacing $2$R$2$W bank with $1$R$1$W banks is not trivial. 
Some of the prior works~\cite{ nannipieri2021risc, li2021high, chen2022cfntt, duong2021configurable, xin2020vpqc, banerjee2019sapphire, Roy2014} address this issue by modifying the NTT algorithm itself.
For example, to use $1$R$1$W memory banks for an NTT, Roy et al.~\cite{Roy2014} proposed a memory-efficient NTT algorithm, which we refer to as the NTT\_swap$2$ algorithm.
Their technique rearranges the output of the two subsequent Butterfly operations to prevent bank conflicts (two $1$R$1$W banks).
As a result, it guarantees that the input pair required by the Butterfly operation in the following stage is in distinct memory banks. 

Although using a $1$R$1$W  memory bank reduces memory space by half, there is still scope for improvement.
To further reduce memory area overhead, we suggest replacing two $1$R$1$W banks of size $N/2$ with four $1$ read/write port ($1$RW) banks of size $N/4$. 
This causes newer bank conflicts, which cannot be addressed by using the NTT\_swap$2$ method.
For example, if a bank receives both read and write requests at the same time, we will need an additional write buffer to store the write requests.
Now the write requests must wait in the write buffer until there are no incoming reads before opportunistically writing back the results.
Although using a write buffer is a good way to solve bank conflicts, the size of the write buffer quickly grows.
Our observation is that if there are $N/4$ continuous read and write accesses to the same bank in a given stage, the write buffer must be the same size as the banks ($N/4$) in order to hold all write requests that overlap with read requests to the same bank.
If we were to use the same size buffers as the memory banks, we incur the same memory overhead as the $1R1W$ memory bank, making this solution impractical. 

We propose a method called NTT\_swap$4$ (refer Algorithm~\ref{Algorithm1}) to avoid using these large write buffers. 
NTT\_swap$4$ reorders the output of four successive Butterfly operations, while NTT\_swap$2$ reorders the result of only two Butterfly operations. (see Figure~\ref{fig:accel-swap4-example}).
This is to ensure that not only the two inputs of each Butterfly operation are stored in different banks (like NTT\_swap$2$), but also the inputs of consecutive Butterfly operations are stored in different banks (NTT\_swap$4$).
As the same bank is not repeatedly used in this scenario, the write buffer can immediately write back the outcomes in the subsequent cycle.
Thus, the write buffer can be as small as one element wide ($\log q$) for a memory bank.
We demonstrate NTT\_swap$4$ technique example (for $N=32$) in Figure~\ref{fig:accel-swap4-example}. 
The order of the Butterfly operations is indicated by the numbers (in red) before each pair of cells.
For example, in stage $0$, the first four Butterfly operations access the following pairs: $(a_{0}, a_{1})$, $(a_{2}, a_{3})$, $(a_{4}, a_{5})$, $(a_{6}, a_{7})$. 
However, stage $1$ expects elements in the order of $(a_{0}, a_{2})$, $(a_{4}, a_{6})$, $(a_{1}, a_{3})$, $(a_{5}, a_{7})$.
To prevent successive Butterfly operations in stage $1$ from accessing the same banks for reads and writes, we reorganize stage $0$'s outputs into the order anticipated by stage 1 (refer Algorithm~\ref{Algorithm1} line~\ref{algnl:swap}).
To carry out this reordering, we use a Reordering Unit (RU).\\

\begin{algorithm}[t]
\caption{{\bf NTT\_swap4} \label{Algorithm1}}
\KwIn{Polynomial $a(x) \in Z_q[x]$ in bit-reversed order}
\KwOut{$NTT(a(x))$ in normal order}

    $m = 2$\;
    \For {$(stage = 0; stage<(\log N-1); stage+=1)$}{
        $\omega = 1$;
        $\omega_m = \omega_n^{2^{\log N-1-stage}}$; 
        $upd\_cnt=1$\;
        \For {$(j = 0; j<m*2; j+=4$)} {
            \For {$(k = 0;k<N; k+=m*4)$}{
                i0=[]; i1=[]\;
                \For{$(l=0; l<4; l+=1)$}{
                    \Switch{$l$}{
                        \lCase{$0$}{
                            $idx = j+k$
                        }
                        \lCase{$1$}{
                            $idx = j+k+2$
                        }
                        \lCase{$2$}{
                            $idx = j+k+m*2$
                        }
                        \lCase{$3$}{
                            $idx = j+k+m*2+2$
                        }
                        
                    }
                    $a[idx] = a[idx] + a[idx+1] * \omega \Mod q$\; \nllabel{algnl:bf1}
                    $a[idx+1] = a[idx] - a[idx+1] * \omega \Mod q$\; \nllabel{algnl:bf2}
                    i0.append($idx$); i1.append($idx+1$)\;
                    \uIf{$upd\_cnt == N/(2^{stage+1})$}{
                        $\omega = \omega * \omega_m \Mod q$; 
                        $upd\_cnt = 1$\;
                    }\lElse{
                        $upd\_cnt+=1$
                    }
                }
                $\begin{aligned}
                    (&a[i0[0]], a[i1[0]], a[i0[1]], a[i1[1]], \\
                    &a[i0[2]], a[i1[2]], a[i0[3]], a[i1[3]]) = \\
                    (&a[i0[0]], a[i0[1]], a[i0[2]], a[i0[3]], \\
                    &a[i1[0]], a[i1[1]], a[i1[2]], a[i1[3]]) \; \nllabel{algnl:swap}
                \end{aligned}$
             }
        }


        $m = $ $(m==N/4)$ ? $2$ : $(m*2)$\;
    }
    \For{$(i=0; i<N; i+=1)$}{
        \Comment{Bit manipulation}
        $phy\_addr=\{i[\log N-3:2], i[\log N-1: \log N-2],i[1:0]\}$ \;
        $a\_out[i]= a[phy\_addr]$\;
    }
    return $a\_out$\;
\end{algorithm}

\noindent \textit{Re-ordering Unit (RU):} 
The RU reorders the output generated by the BFU and writes it back into the memory banks.
A small register array that can store up to $8$ pairs of Butterfly outputs and a reordering logic make up the RU.
Reordering logic begins by sequentially writing the two results of a Butterfly operation and their addresses to the register array in each cycle.
Once there are eight elements in the register array or four pairs of BFU outputs, the reordering logic will send out the elements stored in the registers to the corresponding memory bank. 
Both NTT and iNTT operations can be reordered effectively using RU.
The RU will be active only while doing NTT/iNTT computations based on the $mode$ signal (see Figure~\ref{fig:he_bp}).\\

\noindent \textit{Control Unit (CU):}
The CU consists of two components -- the computation controller and the I/O controller.
Based on the current operation (error sampling, NTT/iNTT, modular addition, and multiplication), the computation controller, which is an FSM, chooses the BFU and RU mode signals. 
In addition, it generates the enable signal and read/write addresses for memory bank accesses. 
During NTT/iNTT operation, the computation controller is also in charge of setting up the NTT unit to compute the twiddle factors on-the-fly.
Depending on the type of CPU request received by the accelerator (encryption or decryption), the I/O controller chooses the necessary set of BFU operations. 
Besides, based on the current en/decryption stage, it also configures the DMA unit for the input/output data transfer to/from memory banks.
\subsection{Further Optimizations to NTT}
\label{subsec:optimization}
In this section, we present a technique to parallelize NTT to further improve its performance.
This is due to the fact that the area-efficient NTT design that we discussed above cannot meet the performance requirements of high-end edge devices and several high-speed applications.
Therefore, by parallelizing the NTT computation, we can improve the performance at the cost of area and power overhead. 
We evaluate this performance vs. area/power trade-off to identify the optimal architectures for different design objectives in Section~\ref{sec:evaluation}. 

To improve the performance of NTT computation, we can perform multiple Butterfly operations in parallel. 
Therefore, we propose a scalable parallel implementation of NTT with multiple BFUs.
To support a parallel NTT architecture using multiple BFUs, we need to address two main requirements: 
\begin{enumerate*} 
\item Multi-port memory banks to read/write multiple BFUs' inputs and outputs simultaneously and 
\item On-the-fly computation of multiple twiddle factors to enable multiple Butterfly operations in parallel.
\end{enumerate*}

\subsubsection{Memory Bank Organization for Parallel NTT}
\label{subsec:memorybanks}
Moving to a multi-port memory bank design is not an efficient solution as increasing the number of ports will quadratically increase the memory area overhead~\cite{azad2022race}. 
To reduce the memory area overhead, we still use the $1RW$ memory banks but we linearly increase the number of memory banks as we increase the number of BFUs.
However, we keep the total memory size the same by proportionally decreasing the size of each memory bank. 
With an increase in the number of memory banks, the data access pattern within each stage of the NTT becomes complicated resulting in data dependencies that need to be carefully managed.
Consequently, our proposed memory reorder technique (see Section~\ref{subsec:ntt}) will not work as it is and requires modifications.

We extend our memory reorder scheme to get rid of the memory bank conflicts by reordering the output of $4 \times \#BFUs$ Butterfly operations instead of reordering the output of only four successive Butterfly operations (as proposed in NTT\_swap$4$ technique).
For example, for $N=32$ with 2 parallel BFUs, in stage $0$, the first eight Butterfly operations access the following pairs: $(a_{0}, a_{1})$, $(a_{2}, a_{3})$, $(a_{4}, a_{5})$, $(a_{6}, a_{7})$, $(a_{8}, a_{9})$, $(a_{10}, a_{11})$, $(a_{12}, a_{13})$, $(a_{14}, a_{15})$. 
However, stage $1$ expects elements in the order of $(a_{0}, a_{2})$, $(a_{4}, a_{6})$, $(a_{8}, a_{10})$, $(a_{12}, a_{14})$, $(a_{1}, a_{3})$, $(a_{5}, a_{7})$, $(a_{9}, a_{11})$, $(a_{13}, a_{15})$.
To prevent successive Butterfly operations in stage $1$ from accessing the same banks for reads and writes, we reorganize stage $0$'s outputs into the order anticipated by stage $1$.

\subsubsection{Twiddle Factor Computation for Parallel NTT}
\label{subsec:tf}
As discussed earlier, to minimize the memory area overhead \rise computes the required twiddle factors on-the-fly instead of storing the precomputed values.
However, now as we increase the number of BFUs for the parallel NTT approach, we need to compute many twiddle factors in parallel, thus introducing stalls in the NTT computation pipeline that offsets the performance gains.
The stalls are introduced because \rise's area-efficient design shares the BFU to compute the twiddle factor and to perform the Butterfly operation.
To eliminate pipeline stalls, we introduce a separate modular multiplier to compute twiddle factors in parallel with the Butterfly operations. 
We note that we also need to increase the number of modular multipliers that are used to compute twiddle factors, as we increase the number of BFUs.

\section{Evaluation}
\label{sec:evaluation}

\subsection{Methodology}
For our analysis, we run all edge-side operations on the following systems in bare-metal mode:
\begin{itemize}
    \item Baseline: BP processor executes all the operations from SEAL-Embedded library.
    \item \race~\cite{azad2022race}: In the \race SoC, the hardware accelerator executes the en/decryption operation, while the remaining operations (error sampling and en/decoding) are performed on the BP processor.
    We modified SEAL-Embedded library to invoke calls to en/decryption operations on the accelerator.
    \item \rise: In the \rise SoC, the hardware accelerator performs error sampling and executes the en/decryption operation while the remaining operations (en/decoding) are performed on the BP processor.
    \rise supports a range of parallel BFUs ($1$ to $32$) within a single NTT operation.
    In our evaluation, \rise-1BFU and \rise-MaxBFU correspond to configurations with $1$ BFU and $32$ BFUs, respectively.
\end{itemize}

All three systems, i.e., \baseline, \race, and \rise, make use of a single core BP configuration ($32$~KB each of Icache and Dcache) running at $1$~GHz. 
We implement all three systems in SystemVerilog and simulate them using VCS. 
The hardware implementation is cycle-accurate and captures the nuances of data movement between all parts of the systems. 
For power, performance, and area evaluation, we use GlobalFoundries $12$nm technology.
We synthesize the logic components in \baseline, \race, and \rise using Synopsys Design Compiler, and use memory compiler for designing the SRAM arrays.

\subsection{Performance Results}

We evaluate \rise performance with different numbers of BFUs ($1$ to $32$) for both \messagetociphertext and \ciphertexttomessage conversion operations for a range of scheme parameters. 
\begin{figure}[b!]
\centering
 \includegraphics[width=\linewidth]{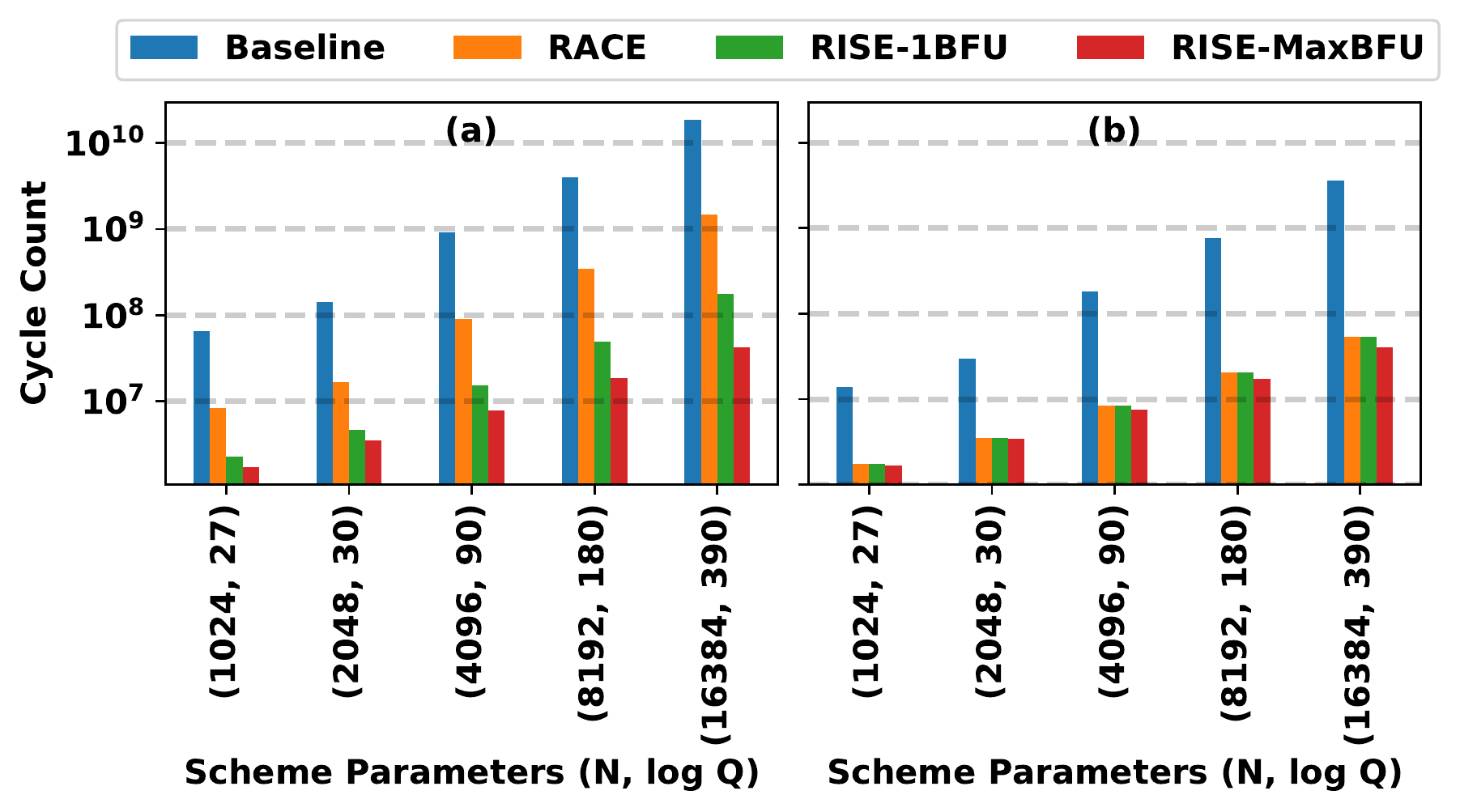}
    \caption{Latency (in clock cycle count) of (a) \messagetociphertext and (b) \ciphertexttomessage conversion operations for \baseline, \race, \rise-$1$BFU, and \rise-MaxBFU with $1$~GHz frequency.}
   \label{fig:end_to_end}
\end{figure}

As shown in Figure~\ref{fig:end_to_end} (a), across different scheme parameter ($N$, $\log Q$) values, \race configuration achieves $7.8\times$-$12.58\times$ and $7.9\times$-$66.08\times$ better performance for \messagetociphertext and \ciphertexttomessage conversion operations, respectively, compared to the \baseline.
The performance improvement in \race is because we offload the encryption and decryption operations to the hardware accelerator, which speeds up encryption by $89.82\times$-$123.76\times$ and decryption by $204.66\times$-$244.44\times$.
In \rise-$1$BFU configuration, on top of encryption and decryption operations, we also offload the error sampling operation to the hardware accelerator, which results in $1726.39\times$-$1734.08\times$ speed up in the error sampling.
However, \rise-$1$BFU configuration achieves just $3.68\times$-$8.29\times$ better performance for \messagetociphertext conversion operation compared to the \race as the performance improvement is limited by Amdahl's law. 

\rise-$1$BFU configuration achieves similar performance as \race for \ciphertexttomessage conversion operation (refer Figure~\ref{fig:end_to_end} (b)), as this conversion does not include the error sampling step. 
As we increase the number of BFUs within the NTT/iNTT operation to perform multiple Butterfly operations in parallel, we observe a speed-up in encryption and decryption operation. 
As shown in Figure~\ref{fig:end_to_end} (a) and (b) for \rise-MaxBFU configuration, with maximum number of BFUs ($32$), compared to \rise-1BFU the \messagetociphertext and \ciphertexttomessage conversion performance improves by $37.7$\%-$414.92$\% and $3.4$\%-$35$\%, respectively. 
Overall, compared to the baseline system, our \race-MaxBFU improves the \messagetociphertext conversion performance by $38.27\times$-$433.14\times$, and the \ciphertexttomessage conversion performance by $8.2\times$-$89.25\times$.
\\

\noindent \textbf{Comparison with Related Works:}
Table~\ref{tab:latency_related_works} presents the performance comparison between \rise and other relevant state-of-the-art works.
The performance comparison includes the latency of NTT operation for a single limb (the dominant operation in \messagetociphertext and \ciphertexttomessage conversion), the SRAM size, the number of memory ports, and the evaluation platform (ASIC or FPGA).
As other existing works use different parameters ($N$ and $\log q$), we compute the performance numbers for \rise for all of these parameter sets.
It is evident from the table that \rise performs faster NTT computation when compared to other designs for all values of $N$ except for \cite{li2021high}.
Moreover, for every value of $N$, \rise utilizes only single-port memory with the smallest SRAM size. 

Thanks to our highly parallel and pipelined NTT computation design that leads to low NTT computation latency.
Li et al.~\cite{li2021high} can perform a single NTT in $38$ cycles as they store precomputed twiddle factors in SRAM, which leads to $10\times$ higher memory requirement than \rise.
In addition, they need dual-port memories to feed the input to their vectorized NTT unit. 
\rise manages to perform NTT computations while using only a single-port memory by leveraging NTT\_swap$4$ method.
Compared to the related work, \rise has the smallest memory footprint because of our on-the-fly twiddle factor generation unit and in-place NTT computation. 
We do not provide a head-to-head comparison of the performance of \rise with the works by~\cite{en_he_fpga, yoon201955nm} as those prior works accelerate en/decryption operations to support HE operations for the BGV scheme while we enable the support for CKKS scheme.

\begin{table*}
\centering
\caption{NTT operation Performance (cycle count) comparison with the state-of-the-art designs in related works. A head-to-head comparison in terms of frequency, power and area numbers cannot be done because of differences in platforms (ASIC vs FPGA) and technology nodes.}
\begin{tabular}{c|c|c|c|cc|c} 
\hline
\multirow{2}{*}{Design} & \multirow{2}{*}{$N$}     & \multirow{2}{*}{$\log q$} & \multirow{2}{*}{\begin{tabular}[c]{@{}c@{}}Latency\\(Clock Cycles)\end{tabular}} & \multicolumn{2}{c|}{SRAM}        & \multirow{2}{*}{Platform}  \\ 
\cline{5-6}
                        &                        &                       &                                                                                  & Size (KB) & R/W Ports            &                            \\ 
\hline
~\cite{nannipieri2021risc}                         & \multirow{6}{*}{256}   & 16                    & 18554                                                                            & ~2.25 KB  & Dual                 & FPGA                       \\
~\cite{banerjee2019sapphire}                        &                        & 24                    & 1289                                                                             & 45 KB     & Single               & ASIC                       \\
~\cite{li2021high}                        &                        & 16                    & 556                                                                              & 13.5 KB   & Dual                 & FPGA                       \\
~\cite{chen2022cfntt}                       &                        & 14                    & 327                                                                              & 22.5 KB   & Dual                 & FPGA                       \\
\rise                    &                        & 30                    & 103                                                                              & 0.93 KB   & Single               & ASIC                       \\
\cite{li2021high}                        &                        & 16                    & 38                                                                               & 10 KB     & Dual                 & FPGA                       \\ 
\hline
~\cite{ye2022pipentt}                       & \multirow{2}{*}{512}   & 16                    & 1074                                                                             & 18 KB     & Dual                 & FPGA                       \\
\rise                    &                        & 30                    & 215                                                                              & 1.87KB    & Single               & ASIC                       \\ 
\hline
~\cite{paludo2022ntt}                         & \multirow{4}{*}{1024}  & 28                    & 2568                                                                             & ~108 KB   & Dual                 & FPGA                       \\
~\cite{ye2022pipentt}                         &                        & 28                    & 2114                                                                             & ~27 KB    & Dual                 & FPGA                       \\
~\cite{su2022highly}                       &                        & 32                    & 650                                                                              & ~355.5 KB & Dual                 & FPGA                       \\
\rise                    &                        & 30                    & 447                                                                              & ~3.75KB   & Single               & ASIC                       \\ 
\hline
~\cite{ye2022pipentt}                       & \multirow{3}{*}{4096}  & 60                    & 8284                                                                             & 110.25 KB & Dual                 & FPGA                       \\
~\cite{su2022highly}                      &                        & 32                    & 3075                                                                             & 355.5 KB  & Dual                 & FPGA                       \\
\rise                    &                        & 30                    & 1918                                                                             & 15KB      & Single               & ASIC                       \\ 
\hline
~\cite{duong2022area}                       & \multirow{2}{*}{16384} & 60                    & 536832                                                                           & 616.5 KB  & Dual                 & FPGA                       \\
\rise                    &                        & 60                    & 34814                                                                            & 480 KB    & Single               & ASIC                       \\ 
\hline
\multicolumn{1}{c}{}    & \multicolumn{1}{c}{}   & \multicolumn{1}{c}{}  & \multicolumn{1}{c}{}                                                             &           & \multicolumn{1}{c}{} &                            \\
\multicolumn{1}{l}{}    & \multicolumn{1}{l}{}   & \multicolumn{1}{l}{}  & \multicolumn{1}{c}{}                                                             &           & \multicolumn{1}{c}{} &                            \\
\multicolumn{1}{l}{}    & \multicolumn{1}{l}{}   & \multicolumn{1}{l}{}  & \multicolumn{1}{c}{}                                                             &           & \multicolumn{1}{c}{} &                           
\end{tabular}
\label{tab:latency_related_works}
\end{table*}

\subsection{Power/Energy Results}
\label{subsec:power_results}
\begin{figure}[t!]
\centering
 \includegraphics[width=\linewidth]{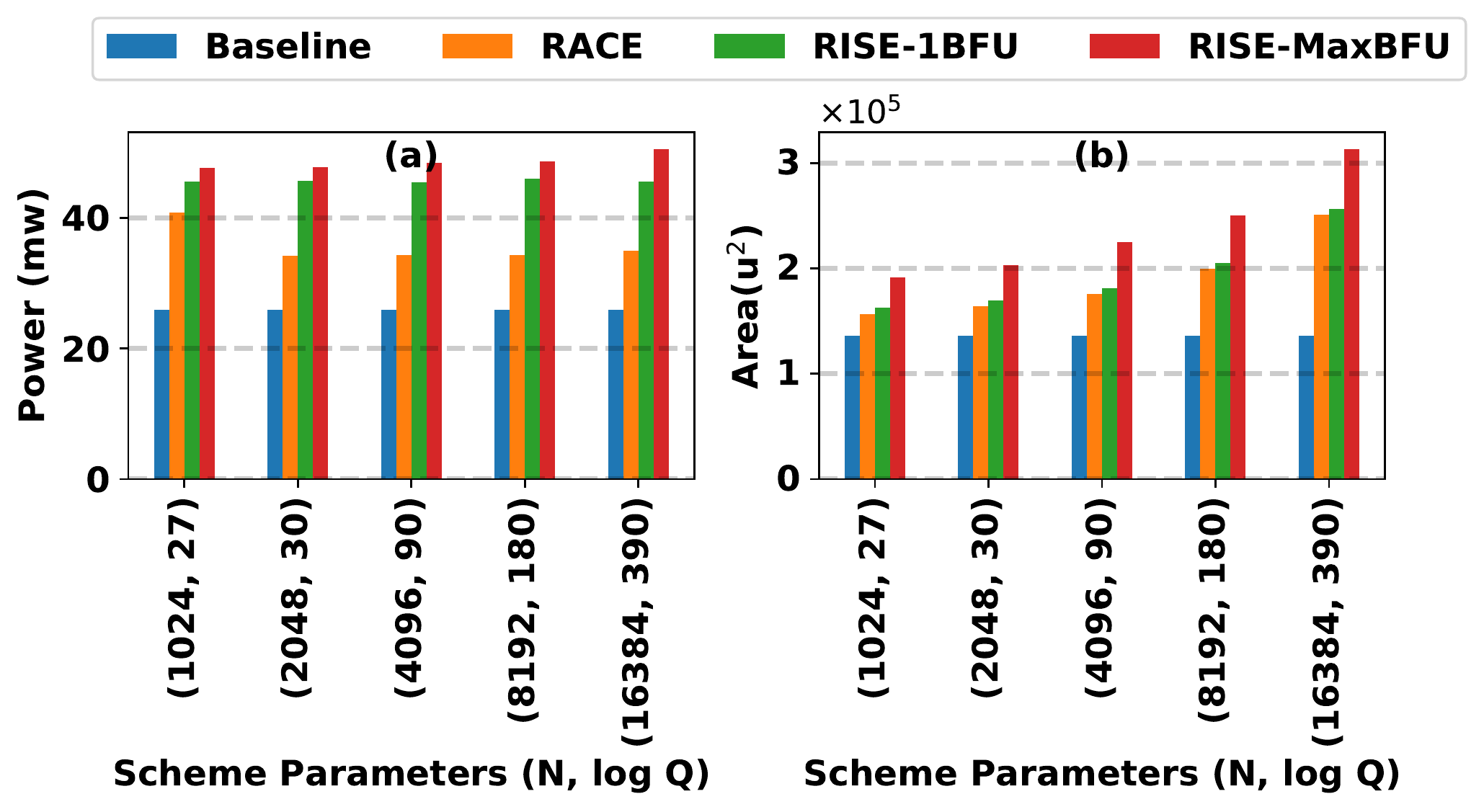}
    \caption{ (a) Power consumption and (b) area utilization for \baseline, RACE, RISE-$1$BFU, and RISE-MaxBFU.}
   \label{fig:area_power_energy}
   \vspace{-0.1in}
\end{figure}
Figure~\ref{fig:area_power_energy} (a) shows the power consumption in the \messagetociphertext and \ciphertexttomessage conversion operations for different scheme parameter ($N$, $\log Q$) values when using \baseline, \race, \rise-$1$BFU, and \rise-MaxBFU systems.
\messagetociphertext and \ciphertexttomessage conversion operations have similar power consumption, within $0.01$\%, and we report the power consumption for the \messagetociphertext\footnote{The \ciphertexttomessage conversion does not perform the error sampling operation and so should have lower power consumption than the \messagetociphertext conversion. However, we did not power gate or clock gate the error sampling unit during the \ciphertexttomessage conversion and so it consumes some power even during the \ciphertexttomessage conversion. 
The error sampling operation takes less than $10$\% of the total time required to perform \messagetociphertext conversion, and so is not the dominant component. 
Hence, the power consumed during \messagetociphertext and \ciphertexttomessage conversion are comparable}.
The total power consumption for a \messagetociphertext and \ciphertexttomessage conversion in the \baseline system is $27.19$~mW, out of which the SRAM power consumption is $41.49\% = 11.4$~mW and the digital logic consumes the remaining power.  
Overall, the power consumption of \race is about $25\%$-$28\%$ (for a range of scheme parameters) higher than the \baseline system for both \messagetociphertext and \ciphertexttomessage conversion operations. 
The increase in the power consumption is due to $41.92\%$-$43.55 \%$ and $3.36\%$-$7.81\%$ power increase in the digital logic and SRAM, respectively.
The power consumption in \rise-$1$BFU configuration increases by $11.62\%$-$30.15$\% compared to \race due to the additional digital logic required for the error sampling unit. 
As we increase the number of BFUs from $1$ to $32$, the power consumption increases by $4.49\%$-$10.98$\% due to the more complex memory banking logic ($14.61\%$-$30.66\%$) and multiple parallel BFUs ($1.01\%$-$4.11\%$).

\subsection{Area Results}
\label{subsec:area_results}
The area of \race is $15\%$ (smallest $N$) to $84\%$ (largest $N$) larger than the area of the \baseline system. (see Figure~\ref{fig:area_power_energy} (b)).
This increase in the area is due to the area required by the accelerator where SRAMs primarily contribute to the increase in area. 
The area overhead in \rise-$1$BFU is ($3.65$\%-$2.14$\%) compared to \race, as error sampling contributes very little to the overall area of \rise.-$1$BFU
With an increase in the number of parallel BFUs, the complexity of control logic and memory banking increases. 
Hence, as shown in Figure~\ref{fig:area_power_energy} (b), by increasing the number of BFUs from $1$ to $32$, the area overhead of \rise-MaxBFU increases by $17.59$\% to $22.38$\% as compared to \rise-$1$BFU for different scheme parameters.

\begin{figure*}[t]
\centering
\includegraphics[width=\linewidth]{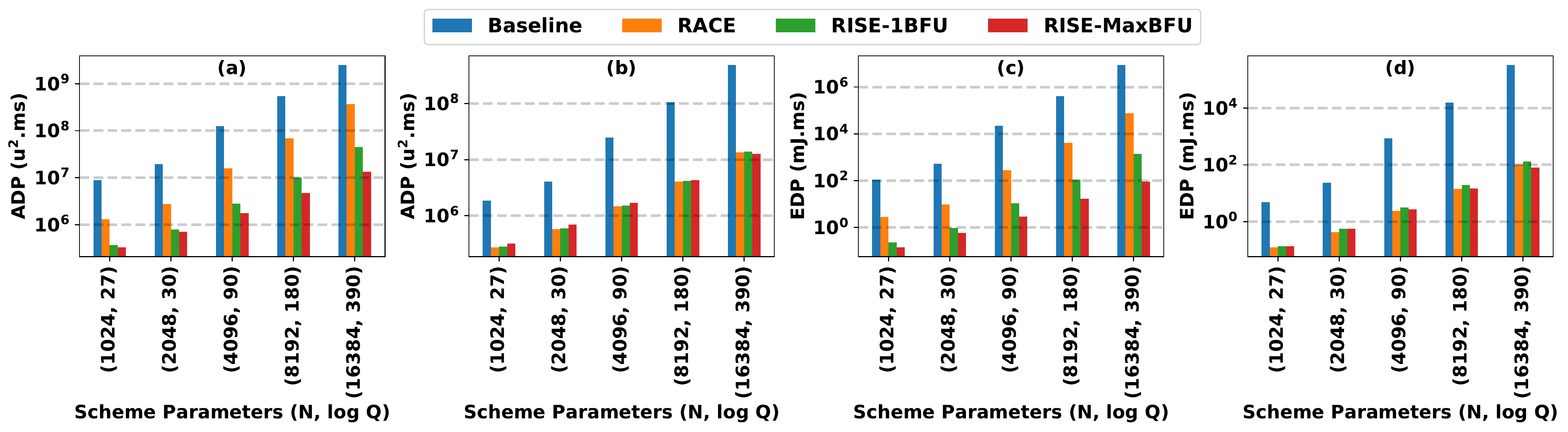}
    \caption{\textit{(a/b) ADP and (c/d) EDP of \messagetociphertext/\ciphertexttomessage conversion operations for \baseline, RACE, RISE-$1$BFU, and RISE-MaxBFU.}}
\label{fig:adp_edp}
\end{figure*}

\subsection{Area and Energy Efficiency}
\label{subsec:area_energy}
\rise aims to improve the performance of \messagetociphertext and \ciphertexttomessage conversion at the cost of increase in area and power.
Thus, Area-Delay Product (ADP) and Energy-Delay Product (EDP) metrics need to be considered for evaluating our \rise design.
Figure~\ref{fig:adp_edp} (a) and (b) compares the ADP value for \baseline, \race, \rise-$1$BFU, and \rise-MaxBFU systems. 
As we can see, \race decreases the \messagetociphertext and \ciphertexttomessage conversion ADP by $6.76\times$-$7.78\times$ and $6.89\times$-$35.80\times$ compared to the \baseline, respectively. 
The improvement is the result of $7.8\times$-$12.58\times$ and $7.95\times$-$66.08\times$ improvement in performance while incurring only a $15$\%-$84\%$ increase in the area. 

The Figure~\ref{fig:adp_edp} (a) also shows that in \rise-$1$BFU, the \messagetociphertext conversion ADP outperforms \race ($3.55\times$-$8.12\times$ lower).
This is due to $3.68\times$-$8.29\times$ performance improvement while incurring only $3.65$\%-$2.14$\% increase in area.
Increasing the number of BFUs from $1$ to $32$ improves the \messagetociphertext conversion ADP by $1.13\times$-$3.39 \times$ for different scheme parameters (due to $1.37\times$-$4.14\times$ performance improvement and $17.59$\%-$22.38$\% area overhead).
For the \ciphertexttomessage conversion the ADP (refer Figure~\ref{fig:adp_edp} (b)) of the \rise-$1$BFU system underperforms \race by $3.65$\% for the smallest $N$ and $2.14$\% for the largest $N$ as there is an increase in area overhead due to the additional error sampling unit, which is not used by \ciphertexttomessage conversion. 
Increasing the number of BFUs to $32$ worsens the \ciphertexttomessage conversion ADP of \rise-$1$BFU by up to $12.3$\% for small $N$ values.
This is because in \rise-$1$BFU the decryption operation only accounts for $3.43$\%-$10.40$\% of the total latency, which when improved by adding parallel BFUs within iNTT, does not improve the performance by the same proportion as the area overhead ($17.59$\%-$22.38$\%).
For large $N$ values, the \ciphertexttomessage conversion ADP increases by up to $10.35$\% as now decryption operation contributes significantly to the total latency, which can be accelerated by instantiating parallel BFUs. 

Figure~\ref{fig:adp_edp} (c) and (d) compare the energy efficiency of \baseline, \race, \rise-$1$BFU, and \rise-MaxBFU systems.
As evident from the figures, EDP follows a similar trend as ADP. 
There is $38.6\times$-$117.09\times$ and $40.19\times$-$3229.81\times$ improvement in the EDP for \messagetociphertext and \ciphertexttomessage conversion, respectively when using \race as compared to the \baseline. 
The EDP of \rise-$1$BFU for \messagetociphertext conversion is $12.19\times$-$52.87\times$ better compared to \race and  by increasing the number of parallel BFUs, EDP further improves by $1.69\times$-$15.51\times$.
Unfortunately, EDP of \rise-$1$BFU for \ciphertexttomessage conversion worsens by $11.62$\%-$30.15$\% compared to \race for the same reason as ADP. 
The EDP of \rise-$1$BFU for \ciphertexttomessage conversion can be improved by clockgating the error sampling unit. 
Moreover, by using 32 BFUs the \ciphertexttomessage conversion EDP improves by $1.71$\%-$64.35$\% compared to \rise-$1$BFU.
This improvement is due to the fact that increasing the number of BFUs improves decryption operation performance, which leads to up to $35$\% performance improvement for \ciphertexttomessage conversion.
We also get up to $21.69$\%\ energy consumption reduction as the BlackParrot core consumes less idle energy.

\subsection{Video Application Evaluation}
\label{subsec:video_app}

\begin{figure}[t!]
\centering
 \includegraphics[width=\linewidth]{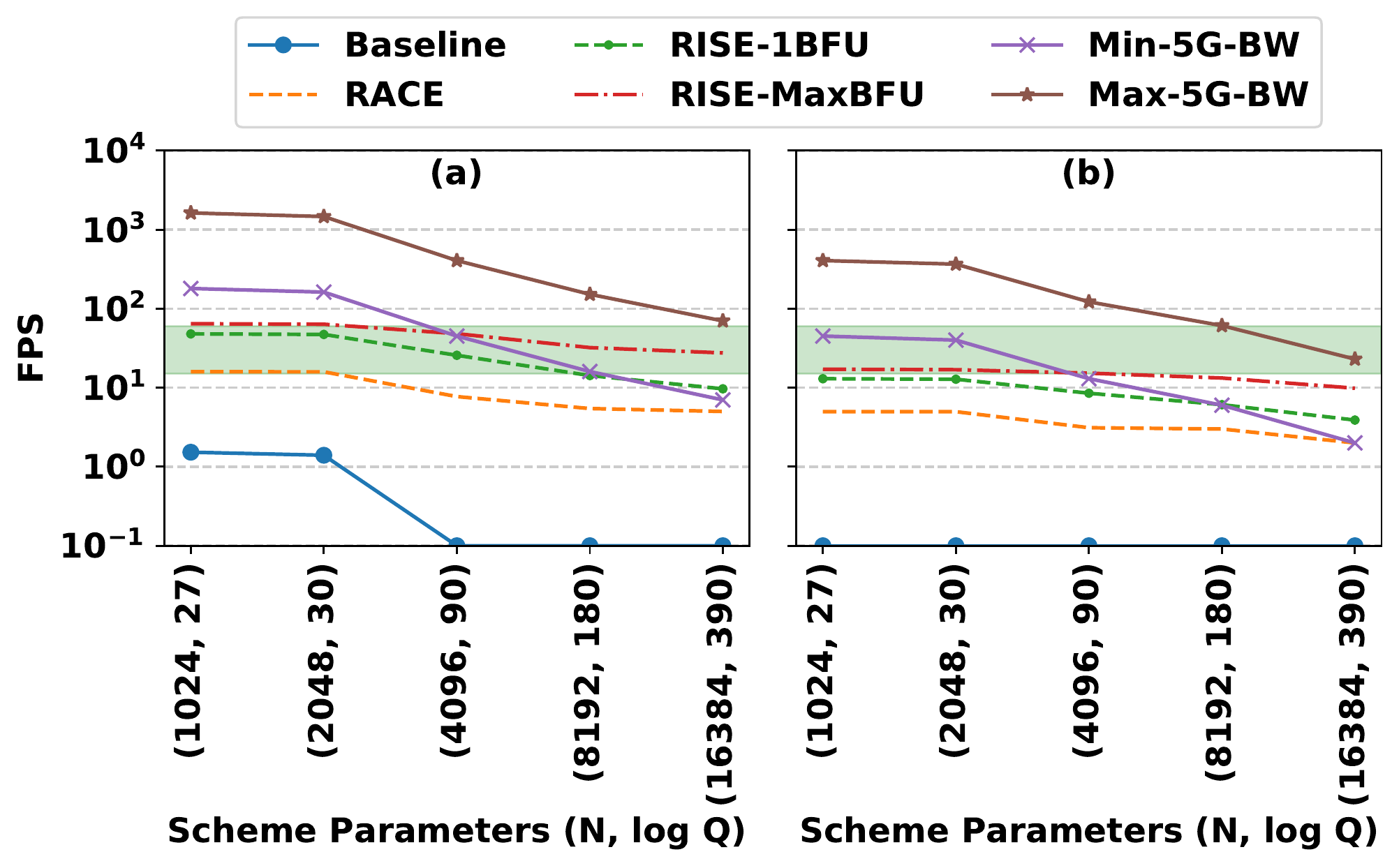}
    \caption{Maximum supported (a) QQVGA and (b) QVGA frame rate per second for mid-band $5$G, \baseline, RACE, RISE-$1$BFU, and RISE-MaxBFU for different $N$ and $\log  Q$ values. The green region indicates the typical frame per second required for surveillance cameras and mobile platforms.}
   \label{fig:fps}
\end{figure}


We evaluate our \rise design using QQVGA and QVGA video frame encryption examples.
For calculating the number of ciphertexts required to encode and encrypt each of these frames refer to Section~\ref{subsec:video_encryption}.
Figure~\ref{fig:fps} (a) and (b) shows the maximum frames per second (FPS) that the \baseline, \race, \rise-1BFU, and \rise-MaxBFU systems can sustain for different scheme parameter ($N$, $\log Q$) values when performing \messagetociphertext conversion operation for QQVGA and QVGA, respectively. 
The frames are sent to the cloud using a mid-band $5$G network, which offers a balance between speed, capacity, and coverage~\cite{8725911}.
As shown in Figure~\ref{fig:fps}, in the regions with maximum bandwidth, mid-band $5$G network can transfer up to $70$ (QQVGA) and $23$ (QVGA) frames per second for the largest $N$ value and in the regions with minimum bandwidth, it can only transfer $7$ (QQVGA) and $2$ (QVGA) frames per second for the smallest $N$ value.
So the throughput of our designs for \messagetociphertext conversion and \ciphertexttomessage conversion should match with these frame rates.

The \baseline system is capable of encrypting up to $2$ QQVGA FPS for $N$ values smaller than $2048$ (refer Figure~\ref{fig:fps} (b)).
However, as we increase $N$ to $4096$ or larger values, it cannot encrypt even a single frame per second. 
On the other hand, for QQVGA, \race encrypts ${\sim}16$ FPS for small values of $N$ and $5$ FPS for the largest $N$ value ($16384$).
So at large values of N we cannot saturate the 5G network at both the maximum bandwidth and minimum bandwidth.

For QVGA resolution, the \baseline system cannot encrypt even one FPS even for the smallest $N$ value ($1024$).
However, \race can encrypt $5$ and $2$ FPS for the smallest and largest $N$ values, respectively.
While \race can support higher FPS than the \baseline, it cannot saturate the 5G network at both the maximum bandwidth and minimum bandwidth for QVGA.

The \rise-$1$BFU system is capable of encrypting up to $48$ QQVGA FPS for $N$ values smaller than $2048$ (refer Figure~\ref{fig:fps}(b)).
For the largest $N$ value, \rise-$1$BFU system is capable of encrypting up to $10$ QQVGA FPS.
As we increase the number of BFUs from $1$ to $32$, the FPS numbers change to $64$ and $27$ QQVGA FPS for the smallest and largest $N$ values, respectively.
Thus, we can saturate the 5G network at minimum bandwidth but not at the maximum bandwidth.

For QVGA resolution (refer Figure~\ref{fig:fps} (a)), \rise-$1$BFU system is capable of encrypting up to $13$ FPS for the smallest $N$ value ($1024$) and $4$ FPS for the largest $N$ value.
The \rise-MaxBFU configuration can encrypt up to $17$ and $10$ QVGA FPS for the smallest and largest $N$ values, respectively.
Thus, we can saturate the 5G network at minimum bandwidth but not at the maximum bandwidth.

Typically surveillance cameras and mobile platforms have an average frame rate of $15$ to $30$ FPS~\cite{usman2018intrusion} (shown by the green highlighted area in Figure~\ref{fig:fps} (a) and (b)). 
For QQVGA resolution, \rise-MaxBFU meets this FPS requirement for all $N$ and $\log Q$ combinations.
For QVGA, \rise-MaxBFU can barely meet the FPS requirement for smaller values of $N$ and $\log Q$.
\section{Conclusion}
\label{sec:conclusion}
In this work, we present \rise, a RISC-V based SoC for \messagetociphertext and \ciphertexttomessage conversion acceleration on the edge to support HE operations in the cloud.
\rise implements several optimizations that enable high performance, and area- and energy-efficient \messagetociphertext and \ciphertexttomessage conversion operations.
These optimizations include data-level parallelism, unified encryption and decryption datapath, memory reuse and memory reordering strategies, and on-the-fly twiddle factor computation.
Our analysis shows that compared to the \baseline and \race, \rise achieves higher performance with lower energy consumption.
As a result, overall \rise is more area and energy efficient than the \baseline and \race system.
Across different $N$ and $\log Q$ parameters, \rise has $471.24\times$-$6191.19 \times$ lower EDP when running a \messagetociphertext  conversion and $36\times$-$2481.44 \times$ lower EDP when running \ciphertexttomessage conversion as compared to \baseline.
Similarly, across different $N$ and $\log Q$ parameters, \rise has $24.06\times$-$55.36 \times$ lower ADP when running a \messagetociphertext conversion and $6.65\times$-$35.05 \times$ lower ADP when running a \ciphertexttomessage conversion as compared to \baseline.




\section{Acknowledgment}
This material is based on research sponsored by Air Force Research Laboratory (AFRL) and Defense Advanced Research Projects Agency (DARPA) under agreement number FA8650-18-2-7856. 
The views and conclusions contained herein are those of the authors and should not be interpreted as necessarily representing the official policies or endorsements, either expressed or implied, of AFRL and DARPA or the U.S. Government.

\bibliographystyle{IEEEtran}
\bibliography{he_tvlsi_2022}

\end{document}